\documentclass[aps,pre,reprint,showpacs,showkeys,amsmath]{revtex4-1}
\usepackage{hyperref}
\usepackage{graphicx}

\begin{document}

\title{Domain-wall complexes in ferromagnetic stripes}
\author{Andrzej Janutka}
\email{Andrzej.Janutka@pwr.wroc.pl}
\affiliation{Institute of Physics, Wroclaw University of Technology, Wybrze\.ze Wyspia\'nskiego 27, 50-370 Wroc{\l}aw, Poland}

\begin{abstract}

Interaction of domain walls (DWs) in ferromagnetic stripes is studied with relevance to the formation of stable complexes of many domains. 
 Two DW system is described with the Landau-Lifshitz-Gilbert equation including regimes of narrow and wide stripes which correspond
 the presence of transverse and vortex DWs. The DWs of both kinds are characterized with their chiralities (the direction 
 of the magnetization rotation in the stripe plane) and polarities (the magnetization orientation in the center of a vortex
 and/or halfvortices), hence, their interactions are analyzed with dependence on these properties. In particular, pairs
 of the DWs of opposite or like chiralities and polarities are investigated as well as pairs of opposite (like) chiralities
 and of like (opposite) polarities. Conditions of the creation of stationary magnetic bubbles built of two interacting DWs
 are formulated with relevance to the situations of presence and absence of the external magnetic field.
\end{abstract}


\maketitle
\newpage

\section{Introduction}

Recent growth of the interest in complexes of domain walls (DWs) in quasi-1D ferromagnets is due to hopes for miniaturization
 of information registers and logical devices based on magnetic nanowires, \cite{par08,hay08,all05,hrk11}, and nanorings \cite{mur09}.
 Currently, the main technological effort is focused on the (straight or curved) magnetic nanostripes whose spin structure 
 is more complex than described within the 1D idealization. It is because nanostructures of the best quality are obtained 
 in this form using lithography methods \cite{kla08}. Mutual interactions of the DWs influence the stability
 of a record of bits encoded in a string of the magnetic domains \cite{hrk11}.
 
In the present paper, I study interactions of DWs in ferromagnetic stripes of nonzero thickness including two regimes 
 of the stripe widths; narrow stripes correspond to the so called transverse DWs while wider stripes to the so called
 vortex DWs \cite{mcm97,che05,min10}. It is shown that both kinds of the DWs are exact stationary 
 solutions to the Landau-Lifshitz-Gilbert (LLG) equation in 2D (different realizations of the cross-tie DW \cite{mid63,uhl09}),
 however, their orientations in the stripe plane are distinguished by different boundary conditions. Both transverse and vortex DWs
 are characterized by two features; a chirality (the direction of magnetization rotation in the stripe plane) which
 is clockwise or counterclockwise one and a polarity (the direction of magnetization alignment in a vortex/halfvortex
 center) which is positive or negative one \cite{mor08,van08}. Pairs of the DWs of the opposite (like) both the chiralities
 and polarities are found to constitute exact stationary solutions to LLG in the absence of external
 field, hence, they do not interact. The interactions of the DWs of opposite chiralities and of like 
 polarities (as well as of like chiralities and of opposite polarities) are studied within a perturbation calculus previously
 developed with relevance to DWs in 1D ferromagnetic wire and in critical systems \cite{jan12}. Evaluating the dependence of energy of
 the two-DW systems on the distance between the DWs, I analyze the creation of stationary bound states of two DWs
 (magnetic bubbles). 

In the presence of an external magnetic field, even DWs of the opposite chiralities and polarities (of like chiralities
 and polarities) interact. The field-induced collision of such DWs is studied applying a method developed to 1D systems
 in \cite{jan11}. For this aim, an extension of the (dissipative) evolution equation of magnetization is performed in a way 
 to describe the dynamics in the limits of large positive and large negative values of time. The collision is found 
 to be accompanied by the reflection of the DWs, hence, by the creation of stable magnetic bubbles.

In section II, the single DW solutions to the LLG equation in magnetic stripes are analyzed. The interactions of the 
 transverse and vortex DWs of opposite chiralities and of like polarities (as well as of like chiralities and of opposite
 polarities) are investigated within the perturbation calculus in section III while the field-induced collisions
 of the DWs of opposite (like) chiralities and polarities are studied in section IV. Main conclusions are summarized in section V.

\section{single DW in ferromagnetic stripe}

I consider stationary DW solutions to the LLG equation in 2D 
\begin{eqnarray}
\frac{\partial{\bf m}}{\partial t}=\frac{J}{M}{\bf m}\times\left(\frac{\partial^{2}{\bf m}}{\partial x^{2}}
+\frac{\partial^{2}{\bf m}}{\partial z^{2}}\right)
+\gamma{\bf m}\times{\bf H}
\nonumber\\
+\frac{\beta_{1}}{M}({\bf m}\cdot\hat{i}){\bf m}\times\hat{i}
-\frac{\alpha}{M}{\bf m}\times\frac{\partial{\bf m}}{\partial t}.
\label{LLG}
\end{eqnarray}
The first term on the r.h.s. of (\ref{LLG}) relates to the exchange interactions while the second (Zeeman) term
 depends on the external magnetic field ${\bf H}=(H_{x},0,0)$, thus, $\gamma$ denotes the gyromagnetic factor.
 The constant $\beta_{1}$ determines strength of the easy-axis anisotropy and $\hat{i}\equiv(1,0,0)$ indicates
 the long axis of the stripe. In nanomagnets with structural disorder (permalloys are the most popular materials
 for nanostripes), the effective spin anisotropy is dominated by dipolar interactions, hence, it is a shape dependent
 and mainly surface effect. Then, for stripes of big 
 enough cross section, the bulk anisotropy is negligible, $\beta_{1}\ll J/w^{2}$, where $w$ denotes the stripe 
 width. Since (\ref{LLG}) is valid only when the constraint $|{\bf m}|=M$ is satisfied, one writes
 equations of the unconstrained dynamics equivalent to (\ref{LLG}). Introducing $m_{\pm}\equiv m_{y}\pm{\rm i}m_{z}$, 
 one represents the magnetization components using a pair of complex functions $g(x,z,t)$, $f(x,z,t)$. The relation 
 between the primary and secondary dynamical variables  
\begin{eqnarray}
m_{+}=\frac{2M}{f^{*}/g+g^{*}/f},
\hspace*{2em}
m_{x}=M\frac{f^{*}/g-g^{*}/f}{f^{*}/g+g^{*}/f}
\label{transform}
\end{eqnarray}
ensures that $|{\bf m}|=M$. Inserting (\ref{transform}) into (\ref{LLG}) leads,
 following the Hirota method for solving nonlinear equations \cite{bog80,kos90}, to 
\begin{eqnarray}
\left[-{\rm i}D_{t}+J(D_{x}^{2}+D_{z}^{2})+\alpha D_{t}\right]&&f^{*}\cdot g
\nonumber\\
-&&\left(\gamma H_{x}+\beta_{1}\right)f^{*}g=0,
\nonumber\\
\left[-{\rm i}D_{t}-J(D_{x}^{2}+D_{z}^{2})+\alpha D_{t}\right]&&f^{*}\cdot g
\nonumber\\
+&&\left(-\gamma H_{x}+\beta_{1}\right)f^{*}g=0,
\label{secondary-eq}\\
(D_{x}^{2}+D_{z}^{2})g\cdot g=0,\hspace*{3em}&&(D_{x}^{2}+D_{z}^{2})f^{*}\cdot f^{*}=0,
\nonumber
\end{eqnarray}
where $D_{t}$, $D_{x}$, $D_{z}$ denote Hirota operators of differentiation 
\begin{eqnarray}
D_{x}^{n}&&b(x,z,t)\cdot c(x,z,t)\equiv
\nonumber\\
&&(\partial/\partial x-\partial/\partial x^{'})^{n}b(x,z,t)c(x^{'},z^{'},t^{'})|_{x=x^{'},z=z^{'},t=t^{'}}.
\nonumber
\end{eqnarray}
For ${\bf H}=0$, stationary single-DW solutions to (\ref{secondary-eq}) take the form
\begin{eqnarray}
f=1,\hspace*{2em}g=u{\rm e}^{kx+qz},
\label{solution_1}
\end{eqnarray}
where 
\begin{eqnarray} 
k^{2}+q^{2}=\frac{\beta_{1}}{J}
\label{condition_1}
\end{eqnarray}
and ${\rm Re}k\neq 0$. I denote $k\equiv k^{'}+{\rm i}k^{''}$, $q\equiv q^{'}+{\rm i}q^{''}$.
 Assuming one of the DW ends to be centered at $x=0$, (then $u={\rm e}^{{\rm i}\varphi}$), the relevant magnetization
 profile [the single-DW solution to (\ref{LLG})] is written explicitly with 
\begin{eqnarray} 
m_{+}(x,z)&=&M{\rm e}^{{\rm i}(\varphi+k^{''}x+q^{''}z)}{\rm sech}[k^{'}x+q^{'}z],
\nonumber\\
m_{x}(x,z)&=&-M{\rm tanh}[k^{'}x+q^{'}z].
\label{profile1}
\end{eqnarray}
I assume $q^{''}\neq 0$ since, in the case $q^{''}=0$, the DW states are similar to DWs in 1D and cannot exist in absence
 of the bulk anisotropy \cite{jan12,jan11,kos90}. Defining $\theta\equiv{\rm arctan}(q^{'}/k^{'})$, following (\ref{condition_1}),
 one finds $k^{''}=-q^{''}\tan(\theta)$ and $k^{'2}-q^{''2}=\beta_{1}/\{J[1+\tan^{2}(\theta)]\}$. Also, I assume the magnetization
 alignments on both the stripe edges to be similar, the ordering is symmetric with respect to the line $z=w/2$. It leads to the
 condition $k^{''}(-wq^{'}/k^{'})+q^{''}w=n\pi$ where $n=1,2,\ldots$ and, finally, to $q^{''}=n\pi/\{w[1+\tan^{2}(\theta)]\}$. 

\begin{figure}
\unitlength 1mm
\begin{center}
\begin{picture}(175,143)
\put(0,-5){\resizebox{75mm}{!}{\includegraphics{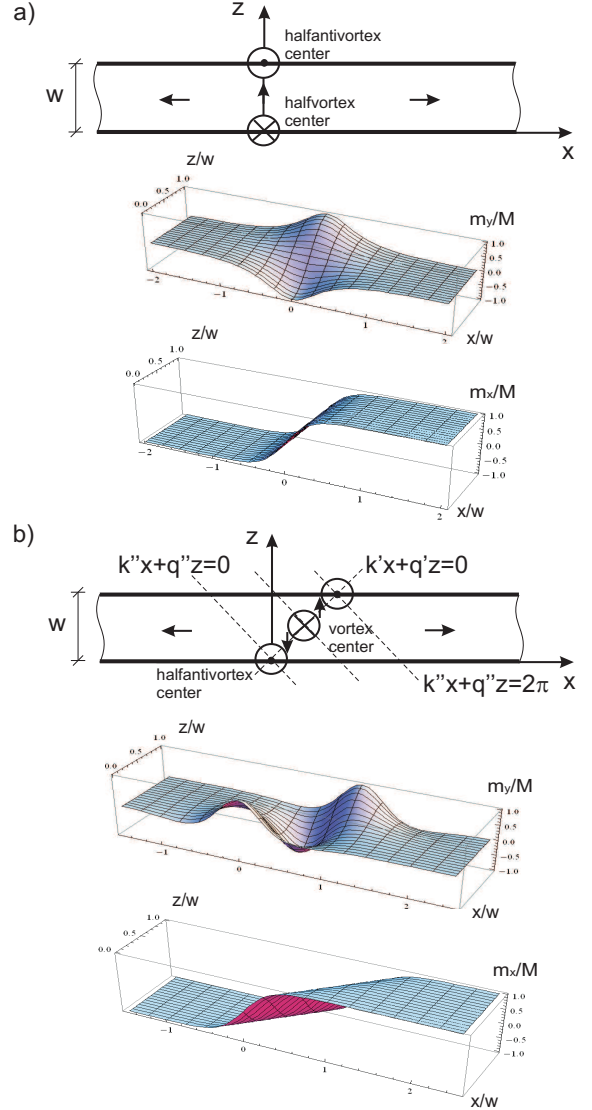}}}
\end{picture}
\end{center}
\caption{DW configurations; a) a transverse DW, b) a vortex DW. In the upper draws, arrows indicate magnetization alignment.}
\end{figure}

Additional boundary condition is related to minimization of the surface (magnetostatic) energy and 
 it discriminates between different values of $\varphi$, $n$, and $\theta$.
 I evaluate the dependence of the energy of the DW on these parameters using 
 the Hamiltonian ${\cal H}={\cal H}_{0}+{\cal H}_{Z}$, where
\begin{eqnarray}
{\cal H}_{0}&=&\frac{J}{2M}\left(\bigg|\frac{\partial{\bf m}}{\partial x}\bigg|^{2}
+\bigg|\frac{\partial{\bf m}}{\partial z}\bigg|^{2}\right)
+\frac{\beta_{1}}{2M}\left[M^{2}-({\bf m}\cdot\hat{i})^{2}\right],
\nonumber\\
{\cal H}_{Z}&=&-\gamma{\bf H}\cdot{\bf m},
\label{hamiltonian}
\end{eqnarray}
(${\cal H}_{Z}$ denotes its Zeeman part). Total energy of the DW $E=E_{0}+E_{Z}+E_{B}$ is the sum 
 of the bulk energy $E_{0}+E_{Z}$, defined by 
 $E_{0(Z)}\equiv\int_{-\infty}^{\infty}\int_{0}^{w}{\cal H}_{0(Z)}{\rm d}z{\rm d}x$, 
 and of the boundary energy $E_{B}$. The contribution $E_{B}$ is determined referring to
 a theorem by Carbou who proved that the magnetostatic energy of any ferromagnetic element of finite 
 thickness $\tau$; $1/\lambda^{2}\int_{S}({\bf m}\cdot{\bf n})^{2}{\rm d}s$ tends to 
 $1/\Lambda_{2}\int_{\partial S}({\bf m}\cdot{\bf n}^{'})^{2}{\rm d}l$ with $\tau\to 0$, \cite{car01}. Here, 
 $S$ denotes the surface of the bulk ferromagnet and $\partial S$ denotes the boundary of the base of its solid,
 ${\bf n}$ is normal to the magnet surface, ${\bf n}^{'}$ denotes the unitary vector outward
 to the line of the base boundary. The coefficient $\Lambda_{2}$ scales with $\tau$ and with a diameter $w$ following
 $\Lambda_{2}\sim\lambda^{2}/[\tau|\log(\tau/w)|]$, (with relevance to a stripe, $w$ represents
 its width \cite{che05,che08}). Writing the formula for $E_{B}$, one has to notice that,
 the Carbou theorem is not strictly applicable to systems with open boundaries, (infinite stripes), however, it
 shows some tendency in ordering at the stripe edges. In particular, it indicates that the magnetostatic 
 interactions in flat magnets induce more than one hard directions of magnetization parallel to the main plane.
 I propose to effectively describe the magnetostatic energy in the form of an integral over the stripe edge 
\begin{eqnarray}
E_{B}=\int_{-\infty}^{\infty}\left[-\frac{2}{\Lambda_{1}}(M^{2}-m_{x}^{2})
+\frac{2}{\Lambda_{2}}m_{z}^{2}\right]_{z=0}{\rm d}x,
\label{E_B}
\end{eqnarray}
where, in contrast to \cite{che08}, the effect of anisotropy relating to the long axis of the stripe is included 
 via the term depending on $\Lambda_{1}$. By analogy to systems with boundaries of finite length that satisfy the Carbou
 theorem, this coefficient is expected to scale with the stripe width following $\Lambda_{1}\sim\lambda^{2}/w$.
 More detailed estimation of $E_{B}$ is performed in Appendix A, (see also \cite{jan12b}).
 
Inserting (\ref{profile1}) into the Hamiltonian (\ref{hamiltonian}), one arrives at 
\begin{eqnarray}
E_{0}(\theta,n)=2JMw
\sqrt{\frac{\beta_{1}}{J}[1+{\rm tan}^{2}(\theta)]+\frac{\pi^{2}n^{2}}{w^{2}}}.
\label{E_0}
\end{eqnarray} 
Estimating $E_{B}$, I divide it into two parts $E_{B}=E_{B1}+E_{B2}$; 
\begin{eqnarray}
E_{B1}(\theta,n)&\equiv&-\frac{2M^{2}}{\Lambda_{1}}\int_{-\infty}^{\infty}
{\rm sech}^{2}\left\{\frac{n\pi x}{w[1+\tan^{2}(\theta)]}\right\}{\rm d}x,
\nonumber\\
E_{B2}(\varphi,\theta,n)&\equiv&\frac{2M^{2}}{\Lambda_{2}}\int_{-\infty}^{\infty}
{\rm sin}^{2}\left\{\varphi-\frac{n\pi\tan(\theta)x}{w[1+\tan^{2}(\theta)]}\right\}
\nonumber\\
&&\times{\rm sech}^{2}\left\{\frac{n\pi x}{w[1+\tan^{2}(\theta)]}\right\}{\rm d}x.
\label{boundary_energy}
\end{eqnarray}
The minimization of $E_{B}(\varphi,\theta,n)$ leads to $\varphi=0,\pi$ independently of other parameters of the DW.  
 It corresponds to the presence of halfvortices (halfantivortices) at the stripe edges (as shown in Fig. 1).
 In the regime of narrow stripes $w/\tau\sim 1$, $\Lambda_{2}\ll\Lambda_{1}$, ($1/\Lambda_{2}$ is small, thus,
 $E_{0}$ dominates over $E_{B}$), via minimization of $E_{0}(\theta,n)$ the smallest possible value of $n$
 is preferable while the minimization of both the energy components $E_{0}(\theta,n)$, $E_{B}(0,\theta,n)$ leads to
 the preferred values of the angle $\theta=0,\pi$. Increase of the stripe width $w$ with fixed thickness $\tau$ results 
 in the transition to the regime $\Lambda_{2}>\Lambda_{1}$. Furthermore, as mentioned previously, for soft-magnetic alloys, 
 $\beta_{1}$ is negligibly small, $\beta_{1}\ll J/w^{2}$, thus, from (\ref{E_0}), $E_{0}(\theta,n)=E_{0}(n)$ 
 is independent of $w$, and $E_{B}$ becomes comparable to $E_{0}$. For this case, via the minimization of $E_{B1}(\theta,n)$,
 the biggest possible value of $\tan^{2}(\theta)$ and the smallest value of $n$ are preferable, whereas, for $\theta\neq 0,\pi$,
 the minimization of $E_{B2}(0,\theta,n)$ indicates big values of $n$ to be preferable. Hence, the transition from the DW state
 of $n=1$, $\theta=0,\pi$ to a state of $n=2$, $\theta\neq 0$ takes place with increase of $w$. For $n=2$,
 the condition $\tan^{2}(\theta)>1$ has to be satisfied in order to $E_{B}(0,\theta,2)<E_{B}(0,0,1)$.

The state $n=1$ and $\theta=0,\pi$ corresponds to $k^{''}=q^{'}=0$, $|q^{''}|=\pi/w$, $|k^{'}|=\sqrt{\pi^2/w^2+\beta_{1}/J}$
 and it is called a {\it transverse DW}. With relevance to the state $n=2$, I take $\beta_{1}=0$, (the shape anisotropy 
 does not affect the dynamical equation while it is completely included in the boundary condition, the minimization
 of $E_{B}$), and $|\theta|$ to be close to its infimum $|\theta|\approx\pi/4$. The resulting magnetization structure
 corresponds to $|q^{'}|\approx|q^{''}|\approx|k^{'}|\approx|k^{''}|\approx\pi/w$ and one calls it a {\it vortex DW}.
 The polarity of vortex (transverse) DW, (the magnetization orientation in the center of vortex/halfvortex, parallel
 or antiparallel to $y$ axis), is determined by value of $\varphi$ while ${\rm sign}(q^{''})$ determines its chirality,
 (the magnetization rotation in the stripe plane, clockwise or anticlockwise to $y$ axis). 
 
\section{Interaction of static domain walls}

Let me analyze the interaction of transverse (vortex) DWs within a perturbation calculus
 developed earlier with relevance to the interactions of DWs in 1D ferromagnet \cite{jan12}. In the present section, 
 I focus on the systems whose interacting DWs are of opposite chiralities and of like polarities
 (or of like chiralities and of opposite polarities) visualized in Fig. 2. I refer to the remaining cases 
 (the DWs of opposite chiralities and polarities as well as the DWs of like chiralities and polarities) in the next section. 
 
\begin{figure}
\unitlength 1mm
\begin{center}
\begin{picture}(175,24)
\put(0,-5){\resizebox{85mm}{!}{\includegraphics{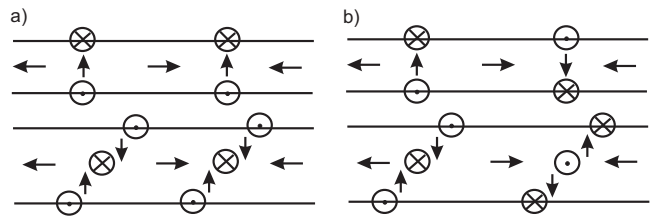}}}
\end{picture}
\end{center}
\caption{Scheme of the magnetization layout in double-DW configurations; a) pair od transverse (parallel vortex) DWs of opposite chiralities
and of like polarities, b) pair od transverse (parallel vortex) DWs of like chiralities and of opposite polarities.}
\end{figure}

Locally, in the vicinity of the center of $j$th DW, ($j=1,2$), one can write the magnetization in the form
\begin{eqnarray} 
{\bf m}(x,z,0)={\bf m}^{(j)}(x,z)+\delta{\bf m}^{(j)}(x,z),
\end{eqnarray}
where ${\bf m}^{(j)}$ denotes the stationary single-DW solution to ($\ref{LLG}$)
\begin{eqnarray} 
m_{+}^{(j)}(x,z)&=&M{\rm e}^{{\rm i}\{\varphi_{j}+\delta_{j}[k^{''}(x-x_{0j})+q^{''}z]\}}
\nonumber\\
&&\times
{\rm sech}\{\sigma_{j}[k^{'}(x-x_{0j})+q^{'}z]\},
\nonumber\\
m_{x}^{(j)}(x,z)&=&-M{\rm tanh}\{\sigma_{j}[k^{'}(x-x_{0j})+q^{'}z]\}
\label{single_DW_int}
\end{eqnarray}
while $\delta{\bf m}^{(j)}$ denotes a perturbation due the presence of another DW. 
 Here $\delta_{1}=\delta_{2}=\pm 1$, $|\sigma_{j}|=1$, $\sigma_{1}=-\sigma_{2}$, and, 
 for $k^{'}>0$, $\sigma_{j}=1$ corresponds to DW of the head-to-head type while $\sigma_{j}=-1$
 to the tail-to-tail type. The phases satisfy $\varphi_{1}=\varphi_{2}$ or $\varphi_{1}=\varphi_{2}\pm\pi$.
 I refer to the case $\delta_{1}=\delta_{2}$ and $\varphi_{1}=\varphi_{2}$ as to the pair of walls
 of opposite chiralities and of like polarities while to the case $\delta_{1}=\delta_{2}$
 and $\varphi_{1}=\varphi_{2}\pm\pi$ as to the walls of like chiralities and of opposite polarities. In terms of vortex DWs, 
 I focus my attention to the interaction of mutually parallel vortex DWs while taking the single DWs profile
 in the form (\ref{single_DW_int}) excludes its use to non-parallel DWs. The non-parallel
 vortex walls overlap at one edge of the stripe much stronger than at the second edge, thus, an effective description
 of their interaction can be performed with 1D model of Ref. \cite{jan12} applied to the stripe-edge magnetization.
 In order to fulfill the constraint $|{\bf m}|=M$, I take the perturbation of the form
\begin{eqnarray}
\delta{\bf m}^{(j)}&=&\left(\pm\frac{m_{x}^{(k)}}{M}-1\right){\bf m}^{(j)}
\pm\frac{m_{x}^{(j)}}{M}\left(0,m_{y}^{(k)},m_{z}^{(k)}\right)
\nonumber\\
&&\mp\frac{1}{M}\left(m_{y}^{(k)}m_{y}^{(j)}+m_{z}^{(j)}m_{z}^{(k)},0,0\right),
\label{perturbation}
\end{eqnarray}
where $k\neq j$, which leads to 
\begin{eqnarray}
m_{x}&=&\pm\frac{1}{M}\left[m_{x}^{(1)}m_{x}^{(2)}-\frac{1}{2}\left(m_{+}^{(1)}m_{-}^{(2)}+m_{+}^{(2)}m_{-}^{(1)}\right)\right],
\nonumber\\
m_{+}&=&\pm\frac{1}{M}\left(m_{+}^{(1)}m_{x}^{(2)}+m_{+}^{(2)}m_{x}^{(1)}\right).
\label{lowest_order_LLG}
\end{eqnarray} 
The plus and minus relate to the bubble magnetization parallel and antiparallel to $x$-axis, respectively. I stress that
 (\ref{lowest_order_LLG}) is irrelevant to the case of DWs of opposite (like) chiralities and polarities,
 $\delta_{1}=-\delta_{2}$, since, for this case, $|m|=M$ was not satisfied. The perturbation of states of the interacting DWs
 should be small. According to (\ref{perturbation}), this condition demands the interacting DWs to be far-enough from
 each other, (functions of the r.h.s. of (\ref{perturbation}) satisfy $|\pm m_{x}^{(k)}/M-1|,|m_{y}^{(k)}/M|,|m_{z}^{(k)}/M|\ll 1$
 if the distance between walls is larger than their width).  

With relevance to the transverse DWs, insertion of (\ref{lowest_order_LLG}), with $|q^{'}|=|k^{''}|=0$,
 into the Hamiltonian (\ref{hamiltonian}) and integration over the stripe area \{$x\in(-\infty,\infty)$, $z\in[0,w]$\}
 leads to the following dependence of the energy of a DW pair on the distance of separation of the walls
\begin{eqnarray}
E_{0}(a)&=&\frac{M}{2}wJk^{'}I^{\pm}(a,0)
\nonumber\\
&=&\frac{M}{2}w\sqrt{J\beta_{1}+\frac{J^{2}\pi^{2}}{w^{2}}}{\rm csch}^{2\mp 1}(a/2)
\nonumber\\
&&\times{\rm sech}^{2\pm 1}(a/2)[-2a+{\rm sinh}(2a)],
\label{energy_transverse}
\end{eqnarray}
where $a\equiv k^{'}(x_{02}-x_{01})$, and $I^{\pm}(a,\theta)$ represent integrals written in Appendix B. 
 The upper signs correspond to the pair of transverse DWs of opposite chiralities
 and of like polarities, the case $\varphi_{1}=\varphi_{2}$, while the lower signs correspond
 to the pair of transverse DWs of like chiralities and of opposite polarities, the case $\varphi_{1}=\varphi_{2}\pm\pi$. 
 Up to a constant of proportionality, the above energy dependences on $a$ are similar to the ones of the pairs of DWs 
 (of opposite chiralities and of like chiralities, respectively) in 1D ferromagnet \cite{jan12}.

For parallel vortex DWs, inserting (\ref{lowest_order_LLG}), with $|q^{'('')}|=|k^{'('')}|=\pi/w$ and $\beta_{1}=0$,
 into the Hamiltonian and integrating it over the stripe area, one arrives at
\begin{eqnarray}
E_{0}(a)&=&MwJk^{'}I^{\pm}(a,1)
\nonumber\\
&=&\pi MJ
\frac{1}{6}{\rm csch}^{5}(a)\{[36a{\rm cosh}(a)-12a{\rm cosh}(3a)
\nonumber\\
&&-24{\rm sinh}(a)-5{\rm sinh}(3a)+3{\rm sinh}(5a)]
\nonumber\\
&&\mp24\cos(a){\rm sinh}^{2}(a)[-2a+{\rm sinh}(2a)]
\label{energy_vortex}\\
&&-\cos(2a)[24a{\rm cosh}(a)-18{\rm sinh}(a)-2{\rm sinh}(3a)]\},
\nonumber
\end{eqnarray}
where upper signs correspond to the pair of DWs of opposite chiralities and of like polarities while lower signs to the pair
 od DWs of like chiralities and of opposite polarities. The energy dependences on 
 the DW separation distance (\ref{energy_transverse}), (\ref{energy_vortex}) are plotted in Figs. 3a and 3b, respectively. 
 The character of extremum (minimum or maximum) of the function $E_{0}(a)$ at $a=0$ indicates that
 the interaction of the (transverse or vortex) DWs of opposite chiralities and like polarities is attractive when their separation
 distance is short while the interaction of DWs of like chiralities and opposite polarities is repulsive in this case.
 To be precise, here, I call the distance between DWs short when it is close to their width since the perturbation
 calculus is applicable to $|a|>1$.  

In the case of presence of an external longitudinal field $H_{x}\neq 0$, the energy dependence on the distance
 of DW separation $E_{0}(a)+E_{Z}(a)$ deviates from these in Figs. 3a and 3b due to non-zero Zeeman term, however,
 such a deviation is small for $a$ close to zero, similar to one in Fig. 1 of Ref. \cite{jan12}. Moreover, the Zeeman term
 is arbitrarily small when restrict considerations to weak-enough external field. With this restriction, due to attractive interaction,
 an (field-induced) collision of transverse or vortex DWs of opposite chiralities and of like polarities is expected
 to result in their mutual annihilation, according to simulations for the transverse DWs \cite{kun09,dju09}.
 Correspondingly, due to repulsive interaction, the collision of transverse or vortex DWs of like chiralities and of opposite
 polarities leads to their mutual reflection. Both the DW annihilation and nucleation have been observed in ferromagnetic
 nanorigs \cite{rot01,cas03}.

Unlike in the case of transverse DWs, the interaction of vortex DWs changes its character
 from attractive to repulsive (or vice versa) with increasing the distance between the DWs. 
  
\begin{figure}
\unitlength 1mm
\begin{center}
\begin{picture}(175,32)
\put(0,-5){\resizebox{85mm}{!}{\includegraphics{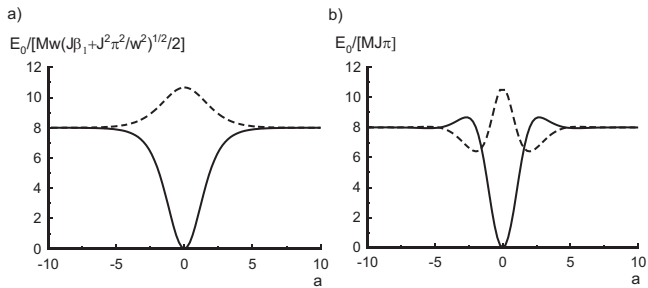}}}
\end{picture}
\end{center}
\caption{Energy of a pair of DWs with dependence on the distance of their separation; a) transverse DWs, 
b) vortex DWs. Solid line - DWs of opposite chiralities and of like polarities, dotted line - DWs of like 
chiralities and of opposite polarities.}
\end{figure}

\section{Field-induced collision of domain walls}

In the present section, the interaction of DWs of opposite chiralities and polarities (as well as of like
 chiralities and polarities) is analyzed. It is shown that such pairs of DWs create static bubbles of magnetization
 which are exact stationary solutions to the LLG equation. Hence, the walls do not interact in absence of any external
 field. The application of a magnetic field in the opposite direction to the bubble magnetization induces motion of both
 the DWs towards each other and, eventually, their collision. It is accompanied by an interaction of the DWs due to a dynamical
 deformation of them which cannot be described in the framework of the above used perturbation calculus. 

Stationary two-DW solutions to (\ref{LLG}) have been found in the absence of anisotropy, for $\beta_{1}=0$.
 Under this condition, for ${\bf H}=0$, inserting the ansatz $f=1$, $g=u{\rm e}^{k_{1}x+q_{1}y}\pm u{\rm e}^{k_{2}x+q_{2}y}$
 into (\ref{secondary-eq}), one finds the relations 
\begin{eqnarray}
k_{1(2)}^{2}+q_{1(2)}^{2}=0,\hspace*{2em}k_{1}k_{2}+q_{1}q_{2}=0. 
\label{opposite-like_conditions}
\end{eqnarray}
Let $k_{j}\equiv k_{j}{'}+{\rm i}k_{j}{''}$, $q_{j}\equiv q_{j}{'}+{\rm i}q_{j}{''}$. For double-DW solutions, 
 $k_{1}^{'}=-k_{2}^{'}=\pm\pi/w$, the above conditions lead to $q_{1}^{''}=-q_{2}^{''}$, $|q_{1(2)}^{''}|=\pi/w$,
 $k_{1(2)}^{''}=q_{1(2)}^{'}=0$ for transverse DWs, and to $k_{1}^{'}k_{2}^{''}=k_{1}^{''}k_{2}^{'}$, 
 $q_{1}^{'}q_{2}^{''}=q_{1}^{''}q_{2}^{'}$, $|k_{1(2)}^{''}|=|q_{1(2)}^{'}|=|q_{1(2)}^{''}|=\pi/w$ for vortex DWs.

Concerning nonstationary DW solutions to the LLG equation for ${\bf H}\neq 0$ and $\alpha\neq 0$, I notice that {\it solitary-wave
 solutions to the LLG equation are relevant in the limit of large positive values of time only, whenever magnetostatic
 effects which lead to the Walker breakdown (an easy-plane anisotropy in the case of 1D ferromagnet) are included} \cite{zha04,tse08,slo73}.
 Hence, studying the dynamics of a pair of DWs which are stationary at the initial moment, one cannot avoid an effect
 of non-adiabatic switching on of the magnetic field. On the other hand, standard approach to the problem of soliton collision
 is based on the analysis of asymptotics of two soliton solutions in the limits $t\to\pm\infty$, e.g. it has been used
 with relevance to collisions of spontaneously propagating topological solitons (DWs) in 1D ferromagnet (in absence of external
 field and dissipation) \cite{bog80}. In the presence of dissipation this method fails since solutions to dissipative equations
 of motion become unphysical in the limit of large negative values of time, because of divergence of energy in this limit.
 In particular, Zeeman energy of any ferromagnetic stripe diverges with $t\to-\infty$ whereas domains magnetized parallel
 (antiparallel) to the magnetic field grow (diminish) with time. In view of the purpose of studying field-induced collisions,
 these facts motivate extension of the dynamical system (\ref{secondary-eq}) within a formalism applicable to the limits of large
 positive and negative values of time. 
 
The method of extension of the dynamical system has been developed in Ref. \cite{jan11}. It is based on 
 a Bateman's idea of doubling the number of freedom degrees when including the dissipation into the standard Lagrange 
 formalism (with relevance to damped harmonic oscillator) \cite{bat31}. I change secondary dynamical equations (\ref{secondary-eq})
 replacing $g$, $g^{*}$, $f$, $f^{*}$ with novel fields of the corresponding set $g_{1}$, $g_{2}^{*}$, $f_{2}$, $f_{1}^{*}$ 
 and of the set of their c.c.s. in a way that the resulting extended system of the equations of motion 
\begin{eqnarray}
\alpha D_{t}f_{1}^{*}\cdot g_{1}-\gamma H_{x}f_{1}^{*}g_{1}={\rm i}D_{t}f_{1}^{*}\cdot g_{1},
\nonumber\\
(D_{x}^{2}+D_{z}^{2})f_{1}^{*}\cdot g_{1}=0,
\label{secondary-eq-pm1}\\
(D_{x}^{2}+D_{z}^{2})g_{1}\cdot g_{1}=0,
\hspace*{1em}(D_{x}^{2}+D_{z}^{2})f_{1}^{*}\cdot f_{1}^{*}=0.
\nonumber
\end{eqnarray} 
and of their conjugates [that differ from (\ref{secondary-eq-pm1}) by the sign of the dissipation constant $\alpha$]
\begin{eqnarray}
-\alpha D_{t}f_{2}^{*}\cdot g_{2}-\gamma H_{x}f_{2}^{*}g_{2}={\rm i}D_{t}f_{2}^{*}\cdot g_{2},
\nonumber\\
(D_{x}^{2}+D_{z}^{2})f_{2}^{*}\cdot g_{2}=0,\label{secondary-eq-pm2}\\
(D_{x}^{2}+D_{z}^{2})g_{2}\cdot g_{2}=0,\hspace*{1em}(D_{x}^{2}+D_{z}^{2})f_{2}^{*}\cdot f_{2}^{*}=0\nonumber
\end{eqnarray}
is, in a formal sense, symmetric with respect to the reversal of the arrow of time, (although, dissipative dynamics 
 of any physical system is irreversible). In the system of eight equations; (\ref{secondary-eq-pm1})-(\ref{secondary-eq-pm2})
 and of their c.c.s, $g_{1}(f_{1})$ is not a c.c. to $g_{2}(f_{2})$, while,
 comparing (\ref{secondary-eq-pm1}) and (\ref{secondary-eq-pm2}), one sees that $g_{2}(x,z,t)$ [$f_{2}(x,z,t)$] 
 can be obtained from $g_{1}(x,z,t)$ [$f_{1}(x,z,t)$] via changing the sign of its parameter $\alpha$.  
 Upon the change $t\to-t$, the system of the novel equations transforms into itself if one accompanies
 this operation by the transform of the novel dynamical variables $g_{1(2)}\to f_{2(1)}$, $f_{1(2)}\to-g_{2(1)}$.
 The above trick is an analogous construction to classical and quantum formalisms for description of dissipative 
 systems in the whole length of the time axis despite the divergence of excitation energy in the limit of large negative
 values of time (non-equilibrium Green functions, thermo-field dynamics, rigged Hilbert space) \cite{kel65}, which
 are all based on the concept of Bateman. The equations (\ref{secondary-eq-pm1}) and their c.c.s., which determine
 the magnetization dynamics for large positive values of time (in particular, for $t\to\infty$), contain the differentials
 of the functions $g_{1}$, $g_{1}^{*}$, $f_{1}$, $f_{1}^{*}$. Therefore, the magnetization vector should be expressed 
 with these functions in the relevant time regime. Writing the magnetization in the form
\begin{eqnarray}
m_{+}=\frac{2M}{f_{1}^{*}/g_{1}+g_{1}^{*}/f_{1}},\hspace{1em}
m_{x}=M\frac{f_{1}^{*}/g_{1}-g_{1}^{*}/f_{1}}{f_{1}^{*}/g_{1}+g_{1}^{*}/f_{1}}
\label{distant-future}
\end{eqnarray} 
ensures that their components satisfy $|{\bf m}|=M$, $m_{x}=m_{x}^{*}$, and they reproduce (\ref{transform}) for $\alpha=0$.
 In the limit $t\to-\infty$, one can analyze the evolution 
 of the magnetization with vector $\tilde{\bf m}$ defined as 
\begin{eqnarray}
\tilde{m}_{+}=-\frac{2M}{f_{2}^{*}/g_{2}+g_{2}^{*}/f_{2}},\hspace{0.2em}
\tilde{m}_{x}=-M\frac{f_{2}^{*}/g_{2}-g_{2}^{*}/f_{2}}{f_{2}^{*}/g_{2}+g_{2}^{*}/f_{2}}.
\label{distant-past}
\end{eqnarray}

The single-DW solution to (\ref{secondary-eq-pm1})-(\ref{secondary-eq-pm2}) is of 
 the form $f^{*}_{2}={\rm e}^{-lt/2}$, $g_{1}=u{\rm e}^{kx+qz-lt/2}$, which leads to the magnetization profile
\begin{eqnarray} 
m_{+}(x,z,t)&=&M{\rm e}^{{\rm i}[\varphi+k^{''}(x-x_{0})+q^{''}z-l^{''}t]}
\nonumber\\
&&\times{\rm sech}[k^{'}(x-x_{0})+q^{'}z-l^{'}t],
\nonumber\\
m_{x}(x,z,t)&=&-M{\rm tanh}[k^{'}(x-x_{0})+q^{'}z-l^{'}t],
\label{22}
\end{eqnarray}
and represents the translationally and rotationally moving transverse DW of $|k^{'}|=|q^{''}|=\pi/w$, $|k^{''}|=|q^{'}|=0$
 or vortex DW of $|k^{'('')}|=|q^{'('')}|=\pi/w$. Here, $l^{'}\equiv{\rm Re}l=\gamma H_{x}\alpha/(1+\alpha^{2})$,
 $l^{''}\equiv{\rm Im}l=\gamma H_{x}/(1+\alpha^{2})$. The nonzero $l^{''}$ corresponds to the DW evolution in strong enough 
 magnetic field that exceeds significantly the Walker-breakdown value $|H_{x}|\gg H_{W}$, \cite{sch74,thi05}. Below the Walker
 breakdown, in a weak external field, the rotation of the magnetization about $x$ axis is suppressed
 by the minimization of surface energy, (a strong effective bi-axial anisotropy \cite{mou07}). The relevant dynamics is described
 with a differing from (\ref{LLG}) primary evolution equation and differing from (\ref{secondary-eq-pm1})-(\ref{secondary-eq-pm2})
 secondary equations which are obtained by exchanging the l.h.s. of (\ref{LLG}) and r.h.s.s. of
 (\ref{secondary-eq-pm1})-(\ref{secondary-eq-pm2}) into zero \cite{jan11}. In this weak-field regime, 
 $l^{'}=\gamma H_{x}/\alpha$, $l^{''}=0$. In the intermediate region, above the Walker breakdown $|H_{x}|>H_{W}$,
 the field-induced motion of the DW is accompanied by a dynamic transformation of its structure \cite{hay08a}.
 The DW oscillatory changes between the transverse and vortex ones \cite{lee07}. My further considerations
 focus on the regime $|H_{x}|<H_{W}$, then the motion of a single DW is a simple translation.

For ${\bf H}\neq 0$, two-DW solution to modified (by taking their l.h.s.s. equal to zero) Eqs. 
 (\ref{secondary-eq-pm1})-(\ref{secondary-eq-pm2}) takes the form 
\begin{eqnarray}
f_{1}^{*}&=&{\rm e}^{\gamma H_{x}t/(2\alpha)},
\nonumber\\
g_{1}&=&u({\rm e}^{k_{1}x+q_{1}z}\pm{\rm e}^{k_{2}x+q_{2}z}){\rm e}^{-\gamma H_{x}t/(2\alpha)},
\nonumber\\
f_{2}^{*}&=&{\rm e}^{-\gamma H_{x}t/(2\alpha)},
\\
g_{2}&=&u({\rm e}^{k_{1}x+q_{1}z}\pm{\rm e}^{k_{2}x+q_{2}z}){\rm e}^{\gamma H_{x}t/(2\alpha)},
\nonumber
\end{eqnarray}
where plus (minus) corresponds to states of a pair of the DWs of like (opposite) chiralities and 
 polarities, while the parameters $k_{1(2)}$, $q_{1(2)}$ satisfy 
 the conditions (\ref{opposite-like_conditions}). Hence, for $k_{1}^{'}=-k_{2}^{'}$ and $q_{1}^{'}=-q_{2}^{'}$, 
 (parallel DWs); $k_{1}^{''}=-k_{2}^{''}$ and $q_{1}^{''}=-q_{2}^{''}$. 
 It should be emphasized that the above single-DW and double-DW solutions to (\ref{secondary-eq-pm1})
 satisfy the original system (\ref{secondary-eq}).
  
Let 
\begin{eqnarray}
\eta_{j}(x,z,t)&\equiv&\sigma_{j}\pi/w(x-x_{0j}+\theta z)-\gamma H_{x}t/\alpha,
\nonumber\\ 
\tilde{\eta}_{j}(x,z,t)&\equiv&\sigma_{j}\pi/w(x-x_{0j}+\theta z)+\gamma H_{x}t/\alpha,
\\
\xi_{j}(x,z)&\equiv&\delta_{j}\pi/w[\theta(x-x_{0j})-z],
\nonumber
\end{eqnarray}
with $\sigma_{1}=-\sigma_{2}=1$, $\delta_{1}=-\delta_{2}$, $|\delta_{1(2)}|=1$, and $\theta=0$ for transverse DWs
 while $\theta=1$ for vortex DWs. For $H_{x}>0$ and $\eta_{k}\ll\eta_{j}\sim 0$, here $k\neq j$, ($k,j=1,2$),
 we find the distant-future limit of the magnetization (\ref{distant-future})
\begin{eqnarray}
\lim_{t\to\infty}m_{+}&=&m_{+}^{(j)}\equiv
(\pm 1)^{j-1}2M\frac{v{\rm e}^{\eta_{j}}{\rm e}^{{\rm i}\xi_{j}}}{1+{\rm e}^{2\eta_{j}}},
\nonumber\\
\lim_{t\to\infty}m_{x}&=&m_{x}^{(j)}\equiv
M\frac{1-{\rm e}^{2\eta_{j}}}{1+{\rm e}^{2\eta_{j}}}.
\label{collision_1-prime}
\end{eqnarray}
Identifying the parameters $x_{0j}$ with the DW-center positions $x_{01}=-x_{02}$,
 [$\ln(u)\propto\sigma_{1}+{\rm i}\delta_{1}\theta=-\sigma_{2}-{\rm i}\delta_{2}\theta$], 
 I introduce the restriction on $v$; $v=1$ or $v=-1$. The magnetization profiles (\ref{collision_1-prime}) 
 describe the motion of well separated DWs of the type presented with (\ref{22}) and correspond to 
 the limit $t\to\infty$ of the DW solutions to the LLG equation in 2D for $|H_{x}|<H_{W}$
 by Slonczewski \cite{slo73,slo74}. 

In the distant-past limit, I describe the magnetization evolution using the field $\tilde{\bf m}$. 
 Following (\ref{distant-past}), for $\tilde{\eta}_{j}\ll\tilde{\eta}_{k}\sim 0$ and $j\neq k$,
\begin{eqnarray}
\lim_{t\to-\infty}\tilde{m}_{+}&=&\tilde{m}_{+}^{(j)}\equiv
-(\pm1)^{k-1}2M\frac{v{\rm e}^{\tilde{\eta}_{k}}{\rm e}^{{\rm i}\xi_{k}}}{1+{\rm e}^{2\tilde{\eta}_{k}}},
\nonumber\\
\lim_{t\to-\infty}\tilde{m}_{x}&=&\tilde{m}_{x}^{(j)}\equiv
-M\frac{1-{\rm e}^{\tilde{2\eta}_{k}}}{1+{\rm e}^{2\tilde{\eta}_{k}}}.
\label{collision_2-prime}
\end{eqnarray}
Considering the collision of DWs which are infinitely distant from each other at the beginning of their evolution, 
 I determine the magnetization dynamic in the limit $t\to-\infty$. 
 via inverting the propagation direction of the kinks of $\tilde{\bf m}$ and reversing 
 the arrow's head of the field vector $\tilde{\bf m}$. Utilizing the properties 
 $\tilde{m}_{+}^{(j)}(x+x_{0k},z,0){\rm e}^{{\rm i}\xi_{k}(-x,-z)}=\tilde{m}_{+}^{(j)}(-x+x_{0k},-z,0){\rm e}^{{\rm i}\xi_{k}(x,z)}$, 
 $\tilde{m}_{x}^{(j)}(x+x_{0k},z,0)=-\tilde{m}_{x}^{(j)}(-x+x_{0k},-z,0)$, for $\eta_{j}\gg\eta_{k}\sim 0$, ($j\neq k$), 
 I arrive at
\begin{subequations}   
\begin{eqnarray}
m_{+}(x,z,t)&=&-\tilde{m}_{+}^{(j)}(-x+2x_{0k},-z,t){\rm e}^{{\rm i}2\xi_{k}(x,z)},
\label{collision_2a}\\
m_{x}(x,z,t)&=&-\tilde{m}_{x}^{(j)}(-x+2x_{0k},-z,t).
\label{collision_2b}
\end{eqnarray}
\end{subequations}

The applicability of the above procedure to the asymptotic evolution of a single DW can be verified
 noticing that any single-DW solution to (\ref{secondary-eq-pm1})-(\ref{secondary-eq-pm2}), [Eq. (\ref{22})], satisfies
\begin{subequations}   
\begin{eqnarray}
m_{+}(x,z,t)&=&-\tilde{m}_{+}(-x+2x_{0},-z,t){\rm e}^{{\rm i}2\xi(x,z)},
\\
m_{x}(x,z,t)&=&\tilde{m}_{x}(-x+2x_{0},-z,t).
\label{collision_2b_plus}
\end{eqnarray}
\end{subequations}
 According to (\ref{collision_1-prime}), (\ref{collision_2a})-(\ref{collision_2b}), two initially closing up DWs of opposite (like)
 chiralities and polarities have to diverge after the collision. 
 The one of the colliding DWs that was initially, for $t\to-\infty$, described with the field ingredient
 $\tilde{m}_{+}^{(j)}$, $\tilde{m}_{x}^{(j)}$, is finally, for $t\to\infty$, described with the field
 ingredient $m_{+}^{(j)}$, $m_{x}^{(j)}$. Hence, during the collision, DWs exchange their parameters
 $x_{01}\leftrightarrow x_{02}$ and their phase factors. The DWs reflect in a way that one can say 
 they pass through each other without changing their velocities and polarities, however, with changing their character
 from the head-to-head one into the tail-to-tail one and vice versa and their chiralities (the process
 is illustrated in Fig. 4). This prediction corresponds to the result of the field-induced collision of a Bloch DW 
 with a Neel DW in 1D ferromagnet, \cite{jan11}, as well as of the collision of spontaneously propagating (in absence
 of the dissipation) topological solitons (DWs) in ferromagnets which are intermediate structures between the Bloch
 and Neel DWs \cite{kos90}.
 
\begin{figure}
\unitlength 1mm
\begin{center}
\begin{picture}(175,54)
\put(0,-5){\resizebox{85mm}{!}{\includegraphics{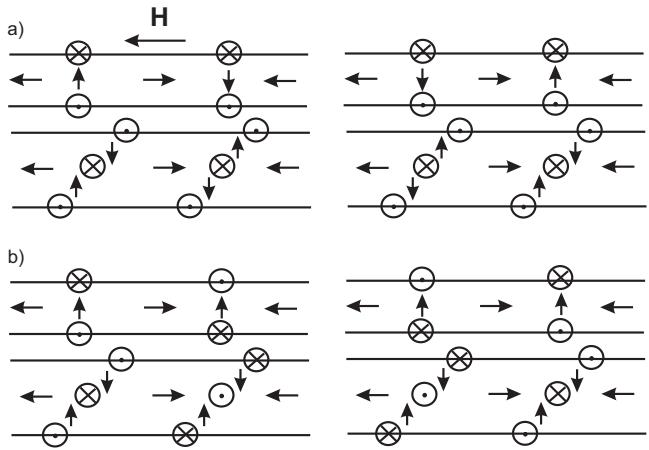}}}
\end{picture}
\end{center}
\caption{Collision of DWs forming a hard magnetization bubble in a stripe. Scheme of the magnetization layout in the initial (final) state 
is shown in the left (right) picture for; a) pairs of transverse and vortex DWs of like chiralities and polarities, 
b) pairs of transverse and vortex DWs of opposite chiralities and polarities.}
\end{figure}

I have considered systems of infinite domains whose energy cannot be defined, however, the smaller 
 a domain is the bigger percentage of the Zeeman part of its energy is lost per time unit due to the DW motion. 
 The field-induced DW reflection induces the motion which contradicts the rule of energy minimization.
 Such a motion has to be decelerated and, eventually, suppressed when the decrease of the DW interaction energy equals
 the increase of the Zeeman energy. It results in formation of a stationary bubble which is a counterpart of hard 
 bubbles in 1D systems \cite{jan11}.

\section{Conclusions}

In terms of the application of multi-DW systems to the magnetic storage, the important result of the present
 study is the prediction of bound states of noninteracting DWs [which are of opposite (like) chiralities
 and polarities] in the absence of any external field. Such bound states are stable 
 with respect to simultaneous change of both the chirality and polarity of one of the DWs of the bubble.
 Unlike stationary bubbles in 1D ferromagnets which are composed of one Neel DW and one Bloch DW, 
 the bubbles in magnetic stripes are composed of DWs of the same energy, hence, they can be stable.
 In 1D systems, the instability is a simple consequence of the fact that the Neel DWs are of higher 
 energy than the Bloch ones and they tend to the reorientation into the Bloch walls in presence 
 of fluctuations \cite{laj79}. The possibility of maintaining stable train of many DWs in magnetic stripes 
 without external power supply (without application of the magnetic field) makes such systems potentially 
 useful as magnetic information registers.   
 
The result of the field-induced collision of DWs of opposite chiralities and polarities (or of like chiralities and polarities)
 is found to be their reflection. Hence, they can form a stripe counterpart of the hard bubbles of magnetization
 of wide ferromagnetic platelets. 
 
Another finding to be stressed is the existence of stationary bound state of two vortex DWs of like chiralities
 and of opposite polarities (a $2\pi$-DW) in the absence of any external field. It follows from Fig. 3b that the energy
 of the relevant pair of vortex DWs $E_{0}(a)$ achieves minims at $a\neq 0$. They correspond to the final state of
 the long-term evolution of a breather. Such a state has no counterpart in 1D ferromagnet while it has a counterpart
 in critical media described with Ginzburg-Landau (or a nonlinear Schrodinger) equation \cite{jan12}.
   
\section*{Acknowledgments}

This work was supported by Polish Government Research Founds for 2010-2012 in the framework of Grant No. N N202 198039.

\appendix
\section{Estimation of boundary energy}

The magnetostatic energy of a magnetic element contains contributions that relate to interactions of surface charges, 
 volume charges, and interaction between surface and volume ones 
\begin{eqnarray}
E_{MS}&=&\int\int\frac{\rho({\bf x})\rho({\bf x}^{'})}{|{\bf x}-{\bf x}^{'}|}{\rm d}V({\bf x}){\rm d}V({\bf x}^{'})
\nonumber\\
&&+\int\int\frac{\sigma({\bf x})\sigma({\bf x}^{'})}{|{\bf x}-{\bf x}^{'}|}{\rm d}S({\bf x}){\rm d}S({\bf x}^{'})
\nonumber\\
&&+\int\int\frac{\sigma({\bf x})\rho({\bf x}^{'})}{|{\bf x}-{\bf x}^{'}|}{\rm d}S({\bf x}){\rm d}V({\bf x}^{'}),
\end{eqnarray}
where $\rho=-\nabla\cdot{\bf m}$, $\sigma={\bf n}\cdot{\bf m}$. Following Ref. \cite{bro65}, reducing one of the spatial
 dimensions with relevance to flat systems of thickness $\tau$ and neglecting volume and base-surface terms, 
 the above expression is transformed into the energy of the boundary of 2D system (up to the multiplayer $\tau$)
\begin{eqnarray}
\tau E_{B}&=&\tau^{2}\int_{\partial S}\int_{\partial S}
\sigma({\bf x})\sigma({\bf x}^{'}){\rm ln}(|{\bf x}-{\bf x}^{'}|/\tau){\rm d}l({\bf x}){\rm d}l({\bf x}^{'})
\nonumber\\
&&+\tau^{2}\int_{\partial S}\int_{S_{base}}
\left[\sigma({\bf x})\rho({\bf x}^{'})+\rho({\bf x})\sigma({\bf x}^{'})\right]
\nonumber\\
&&\times
{\rm ln}(|{\bf x}-{\bf x}^{'}|/\tau)
{\rm d}l({\bf x}){\rm d}S({\bf x}^{'}).
\end{eqnarray}
Here $S_{base}$ denotes the surface of the platelet base.
 For any DW in the stripe, $\rho({\bf x})\sim-\partial m_{x}/\partial x\sim\left[M^{2}-m_{x}^{2}\right]/(M\delta)$
 with $\delta$ denoting the DW width. I estimate the boundary energy with
\begin{eqnarray}
\tau E_{B}&=&\tau^{2}\int_{\partial S}\int_{\partial S}
({\bf n}^{'}\cdot{\bf m})({\bf x})({\bf n}^{'}\cdot{\bf m})({\bf x}^{'}){\rm ln}(|{\bf x}-{\bf x}^{'}|/\tau)
\nonumber\\
&&\times
{\rm d}l({\bf x}){\rm d}l({\bf x}^{'})
-\tau^{2}\int_{\partial S}\int_{S_{base}}
({\bf n}^{'}\cdot{\bf m})({\bf x})\frac{\partial m_{x}}{\partial x}({\bf x}^{'})
\nonumber\\
&&\times
{\rm ln}(|{\bf x}-{\bf x}^{'}|/\tau)
{\rm d}l({\bf x}){\rm d}S({\bf x}^{'})
\nonumber\\
&\sim&
2\delta\tau^{2}{\rm ln}(\delta/\tau)\int_{-\infty}^{\infty}m_{z}^{2}(x,0,0){\rm d}x
\nonumber\\
&&-2aw\tau^{2}{\rm ln}(\delta/\tau)\int_{-\infty}^{\infty}[M^{2}-m_{x}^{2}(x,0,0)]{\rm d}x,
\end{eqnarray}
where $aw$ corresponds to an effective thickness of the surface layer of the stripe edge over which the magnetization 
 is independent on normal coordinate $z$, ($a\ll 1$ and $a\propto\delta$). Since $\delta\sim w$, one arrives at $E_{B}$ of (\ref{E_B}).

\begin{widetext}
\section{Explicit form of integrals}

\begin{eqnarray}
I^{\pm}(a,\theta)&\equiv&\int_{-\infty}^{\infty}\Bigl(\left\{{\rm sech}^{2}(y){\rm tanh}(-y+a)
-{\rm sech}^{2}(-y+a){\rm tanh}(y)\mp\cos(\theta a)\left[-{\rm tanh}(y)+{\rm tanh}(-y+a)\right]
\right.\Bigr.\nonumber\\&&\left.
\times{\rm sech}(y){\rm sech}(-y+a)\right\}^{2}
+\left\{-{\rm sech}(y)\left[{\rm tanh}(y){\rm tanh}(-y+a)+{\rm sech}^{2}(-y+a)\right]
\right.\nonumber\\&&\left.
\pm
\cos(\theta a){\rm sech}(-y+a)\left[{\rm tanh}(y){\rm tanh}(-y+a)+{\rm sech}^{2}(y)\right]\right\}^{2}
\\
&&
+\left\{\sin(\theta a){\rm sech}(-y+a)\left[{\rm tanh}(y){\rm tanh}(-y+a)+{\rm sech}^{2}(y)\right]\right\}^{2}
\nonumber\\
&&\left.
+\left[\sin(\theta a){\rm tanh}(y){\rm sech}(-y+a)\right]^{2}
+\left[{\rm sech}(y){\rm tanh}(-y+a)\pm\cos(\theta a){\rm tanh}(y){\rm sech}(-y+a)\right]^{2}\right){\rm d}y
\nonumber
\end{eqnarray}
\end{widetext}

\end{document}